# Iron Intercalation in Covalent-Organic Frameworks: A Promising Approach for Semiconductors

Srimanta Pakhira,[†,‡] Kevin P. Lucht,[†,‡] and Jose L. Mendoza-Cortes[*,†,‡]

*Department of Chemical & Biomedical Engineering, Florida A&M University - Florida State University, Joint College of Engineering, Tallahassee, Florida, 32310, USA.*

E-mail: mendoza@eng.famu.fsu.edu

Phone: +1-850-410-6298. Fax: +1-850-410-6150

---

[*]To whom correspondence should be addressed
[†]Department of Chemical & Biomedical Engineering, Florida A&M University - Florida State University, Joint College of Engineering, Tallahassee, Florida, 32310, USA.
[‡]Condensed Matter Theory, National High Magnetic Field Laboratory (NHMFL), Scientific Computing Department, Materials Science and Engineering, High Performance Materials Institute (HPMI), Florida State University, Tallahassee, Florida, 32310, USA.

**Abstract**

Covalent-organic frameworks (COFs) are intriguing platforms for designing functional molecular materials. Here, we present a computational study based on van der Waals dispersion-corrected hybrid density functional theory (DFT-D) to design boroxine-linked and triazine-linked COFs intercalated with Fe. Keeping the original $P-6m2$ symmetry of the pristine COF (COF-Fe-0), we have computationally designed seven new COFs by intercalating Fe atoms between two organic layers. The equilibrium structures and electronic properties of both the pristine and Fe-intercalated COF materials are investigated here. We predict that the electronic properties of COFs can be fine tuned by adding Fe atoms between two organic layers in their structures. Our calculations show that these new intercalated-COFs are promising semiconductors. The effect of Fe atoms on the electronic band structures and density of states (DOSs) has also been investigated using the aforementioned DFT-D method. The contribution of the $d$-subshell electron density of the Fe atoms plays an important role in improving the semiconductor properties of these new materials. These intercalated-COFs provide a new strategy to create semi-conducting materials within a rigid porous network in a highly controlled and predictable manner.

# INTRODUCTION

Covalent-organic frameworks (COFs) are an attractive class of crystalline porous materials that can integrate organic units into periodic structures with atomic precision.[1] COFs constitute a class of synthetic materials obtained through mild reversible chemical reactions of organic monomers to form a crystalline framework. COFs can also be obtained through the ordered polymerization of monomers into three-dimensional (3D)[1–3] or two dimensional (2D)[4] crystalline frameworks that present nanometric-sized pores. The COF materials generally contain light elements (such as H, B, C, N, and O) and combine the thermodynamic strength of covalent bonds, such as those found in diamonds and boron carbides, with the functionality of organic units.[5] Thus, COF materials have high structural regularity and diversity, easy modification of frameworks, high porosity, and structural flexibility which make them ideal candidates for various potential applications; particularly in gas adsorption, gas storage/uptake ($H_2$, $CH_4$, etc.),[6–8] as well as energy applications from fuel cells to super capacitors.[9–11]

Present computational and theoretical research show many interesting electronic properties of both the 3D and 2D COF materials with honeycomb-like structures, including: half-metallicity, low intrinsic conductivity, and photoconductivity.[12,13] In addition, due to their $\pi$-stacked layered structures, which can drive the electronic interactions between the nearest neighboring sheets, new properties are expected to emerge from the coupling between individual layers. The low intrinsic conductivity and high charge-carrier mobility in COFs still impose a great challenge regarding their applications in flexible molecular electronics and as thermoelectric materials (which convert heat into electricity or vice versa). A Calcium-intercalated COF (i.e. Ca-COF) was recently proposed by Meng and his co-workers.[14] They computed the equilibrium structures and density of states (DOSs) of the COF using density functional theory (DFT), and showed that this Ca-COF is useful for $H_2$ storage. Very recently, Zhu and Meunier[15] studied the electronic properties of 2D COFs and they showed that their band gap can be fine tuned by appropriate modifications of their structures, specifically by increasing organic chain-links in the framework.

However, instead of modifying the COF structures or increasing the organic linkers in the frameworks, we suggest that the intercalation of transition metal atoms (such as Fe) between the organic layers in COFs can tune the electronic properties by controlling the Fermi energy level ($E_F$) and increasing the number of electron paths (which would maintain the original space group symmetry of the pristine COFs). To the best of our knowledge, the fundamental research into intercalated-COFs has not yet been reported. Thus, works on their intrinsic properties still remain an open opportunity for material discovery. In the present work, we have designed new types of COFs made of boroxine, triazine and benzene rings, and intercalating Fe atoms between two organic layers in the COF. We have also studied their equilibrium structures and electronic properties using first-principles periodic DFT-D[16–22] calculations. The material properties as well as electronic properties were altered by the addition of Fe atoms between two organic layers in the pristine COF, while keeping the same space group symmetry ($P-6m2$) of the pristine COF-Fe-0 (see Scheme 1). The band structures were completely changed and the bands were pushed downword below the Fermi level due to intercalation of five Fe atoms in the pristine COF-Fe-0. In Scheme 1, we showed that the pristine COF-Fe-0, which is an insulator, can be converted into a direct band gap semiconductor by intercalating Fe atoms while keeping the original hexagonal symmetry ($P-6m2$) of the pristine COF. The band gap of the Fe-intercalated COF was reduced by an amount of 1.42 eV from the band gap of the pristine COF-Fe-0. Thus, the intercalation of Fe atoms is a promising approach to changing the electronic properties of a pristine insulating COF into a semiconductor.

Through addition of Fe atoms between layers, seven different types of Fe intercalated-COF materials were created: (i) COF-Fe-1A (one Fe atom was intercalated per unit cell and the Fe atom was placed at the centroid of the boroxine ring), (ii) COF-Fe-1B (one Fe atom was intercalated per unit cell and the Fe atom was placed at the centroid of the triazine ring), (iii) COF-Fe-2 (two Fe atoms were intercalated per unit cell and the two Fe atoms were placed at the centroid of the boroxine and triazine rings), (iv) COF-Fe-3 (three Fe atoms were intercalated per unit cell and they were placed at the centroid of three benzene rings), (v) COF-Fe-4A (four Fe atoms were intercalated per unit cell, whereas three Fe atoms were placed at the centroid of three benzene rings, and one Fe atom was placed at the centroid of the boroxine ring), (vi) COF-Fe-4B (four Fe atoms were intercalated per unit cell, whereas three Fe atoms were placed at the centroid of three benzene rings, and one Fe atom was placed at the centroid of the triazine ring), and (vii) COF-Fe-5 (five Fe atoms were intercalated per unit cell) as shown in Figure 1 and Figure 2, respectively. These materials are thermodynamically stable at room temperature. Thus, this work presents the first comprehensive investigation of bulk crystalline porous COF materials with intercalated Fe atoms. These materials are promising for electronic and nanoporous material applications. In the present article, we have addressed the following points: (a) How to intercalate the Fe atoms between two organic layers of the pristine COF while keeping the original hexagonal symmetry ($P-6m2$) of the pristine COF, (b) The effects of intercalated Fe atoms on the crystal structures and electronic as well as material properties and (c) The role of $d$-subshells electron density of the Fe atoms in improving the semi-conducting properties of the new materials. If these Fe-intercalated COFs can be fabricated, then these materials would have interesting applications in nanoporous materials, electronic devices and thermoelectric materials.

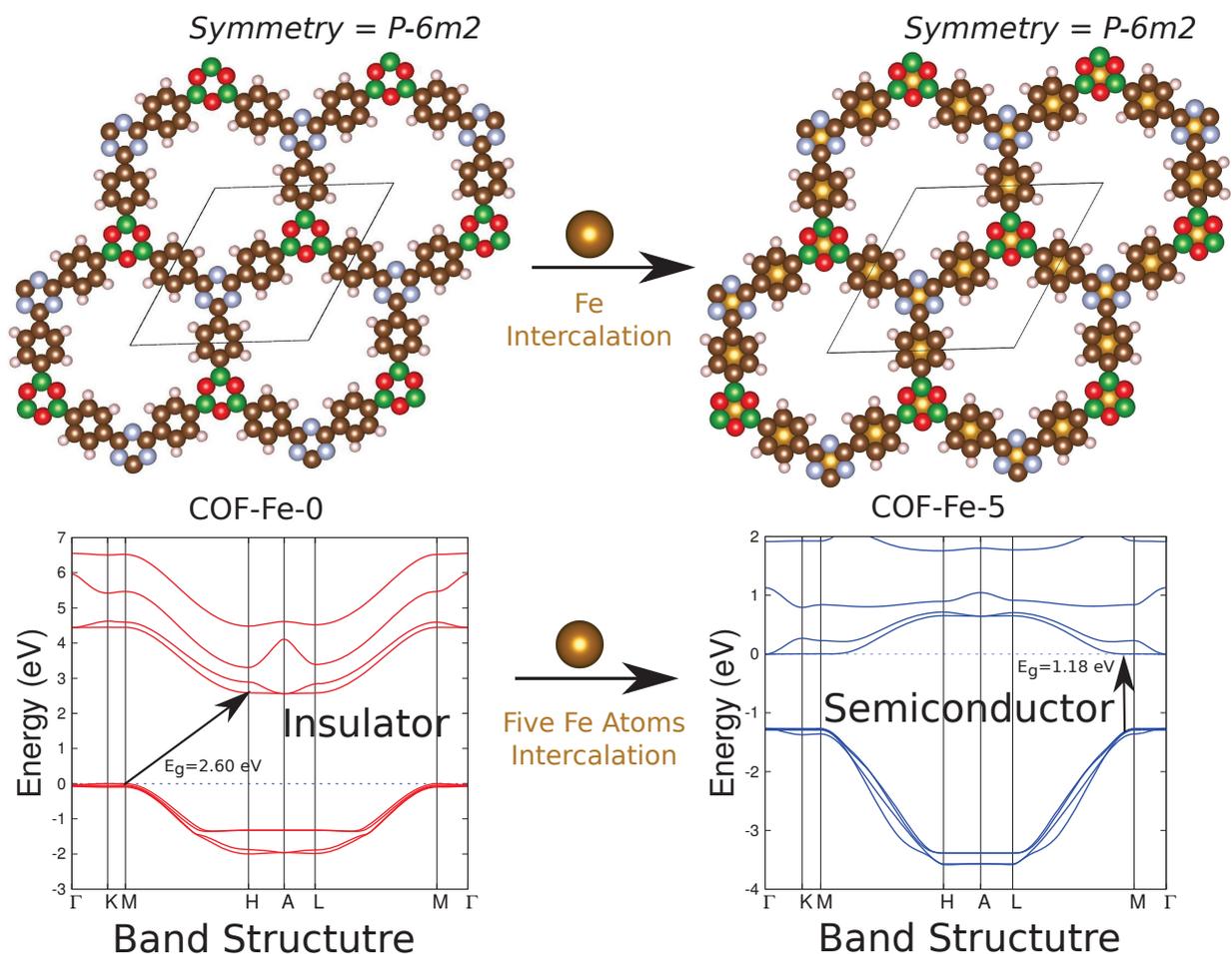

Scheme 1: Optimized crystal structures of the pristine COF-Fe-0 and Fe-intercalated COF-Fe-5 COF materials. The COFs have hexagonal $P-6m2$ symmetry and the unit cells of the COFs are highlighted by a parallelogram. The band structures of both the COFs are shown and the band gap was changed in the Fe-intercalated COF (COF-Fe-5) by intercalating five Fe atoms between two organic layers of the pristine COF (i.e. COF-Fe-0). COF-Fe-5 is a direct band gap semiconductor.

The paper is organized as follows. In Section II, the computational and technical details are briefly introduced. In Section III, we present our results and discussions about the band structures, band gap, density of states, the effects of Fe intercalation and electron density of the $d$-subshells of the transition metals, and the changes of the band gaps. The paper ends with Section IV where a short conclusion is given.

## METHODS AND COMPUTATIONAL DETAILS

First-principles calculations based on hybrid DFT (B3LYP) were used to perform all the periodic boundary computations as implemented in the *ab initio* CRYSTAL14 suite code, which makes use of localized (Gaussian) basis sets.[23] Dispersion effects (DFT-D) have been included;[19] i.e. semi-empirical Grimme's (-D2) dispersion corrections were added in order to incorporate van der Waals (vdW) dispersion interactions in the system.[19,24–28] DFT-D has been shown to give quite accurate thermochemistry for both covalently bonded systems and systems dominated by dispersion forces.[29] However, the dispersion corrections are only available for the first 5 rows of the periodic table. Thus, the geometries of the bulk crystal structure of the aforementioned pristine COF and Fe intercalated-COF materials were opti-mized using the unrestricted dispersion corrected B3LYP hybrid; UB3LYP-D,[16–22] which for simplicity, this is termed DFT-D throughout the text. The unrestricted DFT-D approach was used to consider the spin polarization in the present calculations. Note that this code performs the calculations based on Gaussian basis sets and not on plane waves.[23] Thus, the algorithms are different than other plane-wave based codes. Despite this, all these methods reach similar solutions. During the optimization and single point energy calculations, 'SPIN' and 'SPINLOCK' keywords were used to specify the unrestricted wave functions and the total spin of the intercalated-COF materials, respectively. The 'ATOMSPIN' keyword was also used to specify the individual spins of the Fe atoms in the Fe-intercalated COFs (COF-Fe-1A, COF-Fe-1B, COF-Fe-2, COF-Fe-3, COF-Fe-4A, COF-Fe-4B, and COF-Fe-5; see Figure 1 and 2). In the present computation, triple-zeta valence with polarization quality (TZVP) Gaussian basis sets were used for H, B, C, N, O, and Fe atoms.[30] The threshold used for evaluating the convergence of the energy, forces, and electron density was $10^{-7}$ a.u. for each parameter. The mathematical expression used to calculate the binding energy, $\Delta E$, is given below:

$$\Delta E = E_{COF-Fe-0} - nE_{Fe}$$ ; where n is the number of Fe atoms intercalated in the COFs.

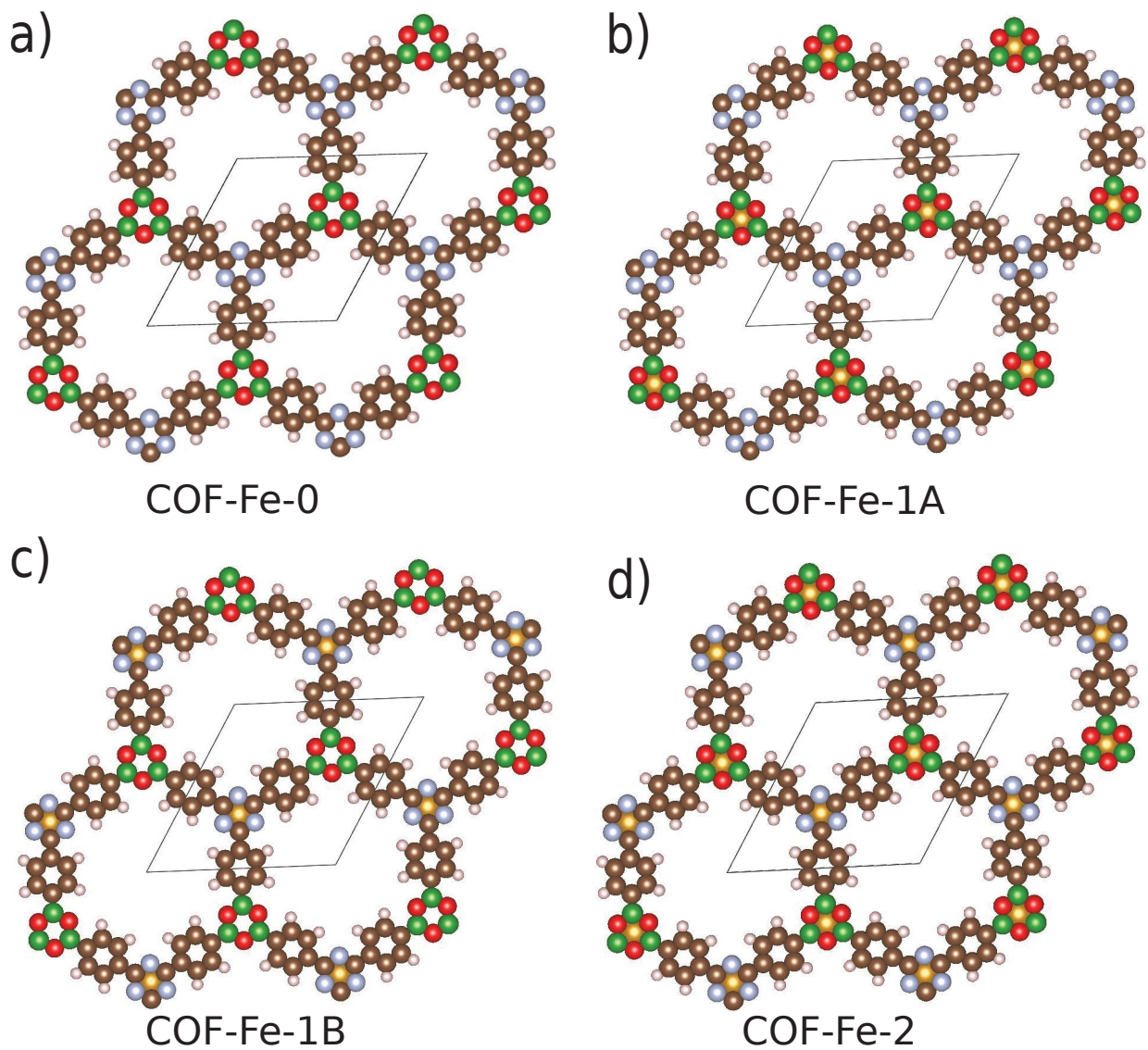

Figure 1: Optimized crystal structures of the pristine (a) COF-Fe-0 and Fe-intercalated (b) COF-Fe-1A, (c) COF-Fe-1B, and (d) COF-Fe-2 COF materials. The COFs have hexagonal $P-6m2$ symmetry and the unit cells of the COFs are highlighted by a parallelogram.

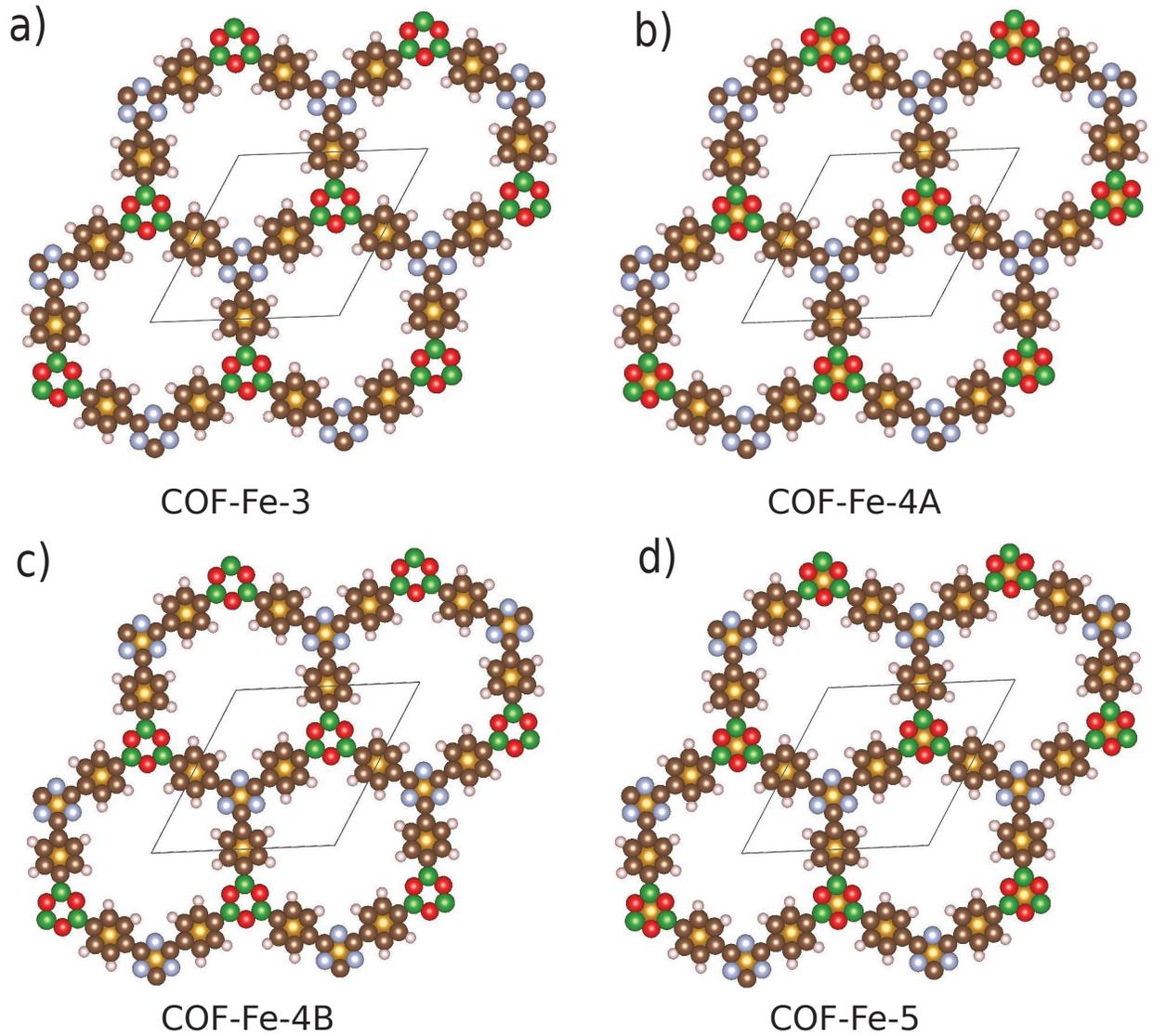

Figure 2: Optimized crystal structures of Fe-intercalated (a) COF-Fe-3, (b) COF-Fe-4A, (c) COF-Fe-4B, and (d) COF-Fe-5 COF materials. The COFs have hexagonal $P-6m2$ symmetry and the unit cells of the COFs are highlighted by a parallelogram.

The electronic properties of these materials were calculated at the optimized equilibrium structures of the pristine COF and intercalated-COF materials using the same level of theory (i.e. UB3LYP-D2). Integrations inside of the first Brillouin zone were sampled on $4 \times 4 \times 16$ Monkhorst-Pack[31] k-mesh grids for all the COFs during geometry optimization, and $20 \times 20 \times 20$ Monkhorst-Pack k-mesh grids for the calculations of band structure and density of states. The reciprocal space for all the structures was sampled by a Γ-centered Monkhorst-Pack scheme with a resolution of around $2\pi$ x $1/60$ Å$^{-1}$. The band pathway followed the symmetry points: $\Gamma - K - M - H - A - L - M - \Gamma$ (Figure 3, 4, 5). The total DOSs were plotted using the atomic orbitals of carbon (C), hydrogen (H), nitrogen (N), boron (B), oxygen (O), and intercalated iron (Fe) atoms. To investigate the contribution of the $d$-subshell electron density of the intercalated Fe atoms in the total DOSs, the DOSs of the $d$-subshells were computed and plotted in Figure 3 - 5. Visualization and analysis were performed using VESTA.[32]

The stability of the pristine and Fe-intercalated COFs was checked by vibrational analysis and by calculating the total energy and binding energy of the systems, for all materials (pristine and Fe-intercalated COFs). None of the systems have negative frequencies, which means that they are thermodynamically stable. The binding energy is negative, which shows that their formation is favorable.

# RESULTS AND DISCUSSIONS

The optimized crystal structures of the pristine COF and Fe-intercalated COFs, COF-Fe-0, COF-Fe-1A, COF-Fe-1B, COF-Fe-2, COF-Fe-3, COF-Fe-4A, COF-Fe-4B, and COF-Fe-5, are shown in Figures 1a, 1b, 1c, 1d, 2a, 2b, 2c, and 2d respectively. We intercalated the Fe atom(s) in order to maintain the original $P-6m2$ space group symmetry of the pristine COF-Fe-0. One Fe atom was intercalated at the centroid of the boroxine ring between two organic layers in one unit cell of the pristine COF in order to form COF-Fe-1A (see Figure 1b). Similarly, one Fe atom was placed at the centroid of triazine ring to form COF-Fe-1B (see Figure 1c). In COF-Fe-2, two Fe atoms were added at the centroid of both the boroxine and triazine rings; see Figure 1d. It is impossible to place Fe atom(s) at the centroid of the benzene rings, since it would break the $P-6m2$ space group symmetry in COF-Fe-1A, COF-Fe-1B and COF-Fe-2. Similarly, three Fe atoms have been placed at the centroid of three benzene rings between two layers in one unit cell of COF-Fe-3. In one case, the fourth Fe atom was placed at the centroid of the boroxine ring in COF-Fe-3 to form COF-Fe-4A (see Figure 2b), and in other case, the fourth Fe atom was placed at the centroid of the trazine ring to form COF-Fe-4B; see Figure 2c. The fifth Fe atom is placed at the centroid of the triazine ring in COF-Fe-4A to form COF-Fe-5 (Figure 2d). The coordinates of the optimized COF's crystal structures are reported in the Supporting Information. The COF materials studied here were energetically stable. All of the COFs presented here were designed as ideal honeycomb-like structures with the **bnn** topology, and the final material consisted of one triazine, one boroxine, and three benzene rings in one unit cell with the $P-6m2$ space group symmetry.

Due to the similar crystal structure but slightly different chemistry, it is interesting to compare the basic electronic structures and properties of the COFs. The equilibrium lattice constants ($a$, $b$, $c$), various average equilibrium bond distances (C-C, B-O, C-B, C-N) and the average distance between two layers i.e. intercalation distance ($d$) are reported in Table 1, while their band gap ($E_g$) and binding energies ($\Delta E$) are shown in Table 2. The lattice constants and various bond distances gradually changed when the Fe atoms were added inside the COFs. Of all the intercalated-COFs studied here, the boroxine and triazine rings are slightly different from the benzene rings. For example, both the boroxine and triazine rings are not perfect hexagon rings in COF-Fe-3, since the angles ∠BOB and ∠NCN are 121.6° and 124°, respectively. This is a small deviation from the expected 120°. This is because the electronegativity of the O and N atoms is stronger (i.e. there is a lone pair of electrons) than that of the B and C atoms, and both the boroxine and triazine rings have less aromatic character compared to pure benzene rings.[33–35] The distances $d$ between two identical organic layers in COF-Fe-0, COF-Fe-1A, COF-Fe-1B, COF-Fe-2, COF-Fe-3, COF-Fe-4A and COF-Fe-4B increased as more Fe atoms were added between the layers (see Table 1) except COF-Fe-5, where the distance $d$ was decreased by 0.002 Å relative to COF-Fe-4B. Thus, adding Fe between two boroxine rings had a different effect than adding Fe between benzene and triazine rings.

The total energy of the systems was computed using the above mentioned DFT-D method; and these values are reported in the Supplementary Information. The present study found that COF-Fe-1B was more stable than COF-Fe-1A by about -35.97 kcal mol$^{-1}$. Similarly,

COF-Fe-4B was more stable than COF-Fe-4A by about -38.22 kcal mol$^{-1}$. The binding energy ($\Delta E$) of each Fe-intercalated COF was calculated with respect to the pristine COF-Fe-0 (see Table 2). The binding energy calculations show that Fe intercalation at the centroid of boroxine rings is less favorable than triazine rings.

Table 1: The lattice constants ($a$, $b$, $c$), average bond distances, and intercalation distance ($d$) between two layers of COFs. The bond distances are reported in Å.

| COFs | $a$ | $b$ | $c$ | C-C | C-B | B-O | C-N | $d$ |
|---|---|---|---|---|---|---|---|---|
| COF-Fe-0 | 14.85 | 14.85 | 3.31 | 1.39 | 1.53 | 1.38 | 1.34 | 3.241 |
| COF-Fe-1A | 14.85 | 14.85 | 3.31 | 1.40 | 1.51 | 1.44 | 1.34 | 3.312 |
| COF-Fe-1B | 14.80 | 14.80 | 3.29 | 1.40 | 1.53 | 1.38 | 1.38 | 3.292 |
| COF-Fe-2 | 14.95 | 14.95 | 3.37 | 1.39 | 1.51 | 1.44 | 1.41 | 3.372 |
| COF-Fe-3 | 14.83 | 14.83 | 3.44 | 1.44 | 1.51 | 1.39 | 1.34 | 3.443 |
| COF-Fe-4A | 14.95 | 14.95 | 3.45 | 1.44 | 1.50 | 1.43 | 1.34 | 3.448 |
| COF-Fe-4B | 14.92 | 14.92 | 3.45 | 1.43 | 1.51 | 1.39 | 1.40 | 3.452 |
| COF-Fe-5 | 15.03 | 15.03 | 3.45 | 1.43 | 1.51 | 1.43 | 1.39 | 3.450 |

Table 2: The band gap energy $E_g$ and binding energy ($\Delta E$) of the pristine COF-Fe-0 and iron intercalated-COFs, COF-Fe-1A, COF-Fe-1B, COF-Fe-2, COF-Fe-3, COF-Fe-4A, COF-Fe-4B, and COF-Fe-5 materials. The band gap energy and binding energy are expressed in eV. Effective mass ($m_e^*$) of the Fe-intercalated COFs is reported here and it is expressed in terms of electron mass.

| COFs | $E_g$ | $\Delta E$ | $m_e^*$ | Material |
|---|---|---|---|---|
| COF-Fe-0 | 2.60 | Reference | – | Insulator |
| COF-Fe-1A | 2.84 | -1.97 | – | Insulator |
| COF-Fe-1B | 1.92 | -3.53 | 0.201 | semiconductor |
| COF-Fe-2 | 1.97 | -4.47 | 0.262 | semiconductor |
| COF-Fe-3 | 1.16 | -14.47 | 0.271 | semiconductor |
| COF-Fe-4A | 1.38 | -16.27 | 0.587 | semiconductor |
| COF-Fe-4B | 1.12 | -17.93 | 1.502 | semiconductor |
| COF-Fe-5 | 1.18 | -19.70 | 0.169 | semiconductor |

After obtaining the optimized geometry of the pristine COF and intercalated-COFs, we investigated their electronic properties. The band structure and density of states (DOSs) of COF-Fe-0, COF-Fe-1A, COF-Fe-1B, COF-Fe-2, COF-Fe-3, COF-Fe-4A and COF-Fe-4B, and COF-Fe-5 were computed and are shown on the right side of Figures 3-5. Additionally, the contribution made toward the electrons density by the *d*-subshells of the Fe atoms in the total DOSs of the Fe-intercalated COFs was computed and denoted by **Fe d** in Figures 3-5. The band structure was plotted along a high symmetric *k*-direction in the first Brillouin zone, given by $\Gamma - K - M - H - A - L - M - \Gamma$. It was found that the pristine bulk COF material, COF-Fe-0, behaves like an insulator (Figure 3a). However, when the pristine COF was intercalated with Fe atoms, the electronic properties are drastically changed. In other words, the band structures and DOSs indicate that COF-Fe-0 has an indirect band gap around 2.60 eV, resulting in an insulator, while COF-Fe-1B is a semiconductor with an indirect band gap of 1.97 eV as shown in Table 2. This band gap is smaller than the values for the pristine ordinary 3D COF materials of 2.3-4.2 eV calculated by Lukose et al.[36] These calculations show that intercalating Fe inside pristine COF can lower the band gap by as much as 0.60 eV, which is remarkable. Interestingly, the present DFT-D calculation found that the band gap was increased by an amount of 0.24 eV when a Fe atom was intercalated at the centroid of the boroxine ring in COF-Fe-1A. This result indicates that boroxine ring is not conducting and it has less aromatic character than the pure benzene ring. The band structures and DOSs of both the COF-Fe-1A and COF-Fe-1B are shown in Figure 3b-c. These calculations show that bands are moved below the Fermi energy level by intercalating the Fe atoms at the centroid of the boroxine or triazine rings. The present DFT-D calculation reveals that the *d*-subshells of the Fe atom provide the electron density in the total DOSs in the COF-Fe-1B, whereas *d*-subshells of the Fe atom have less contribution in the total DOSs in the COF-Fe-1A. The smaller contribution to the DOSs results in an insulator, and indicates that the Fe atom created covalent bonds with the boroxine rings in the framework.

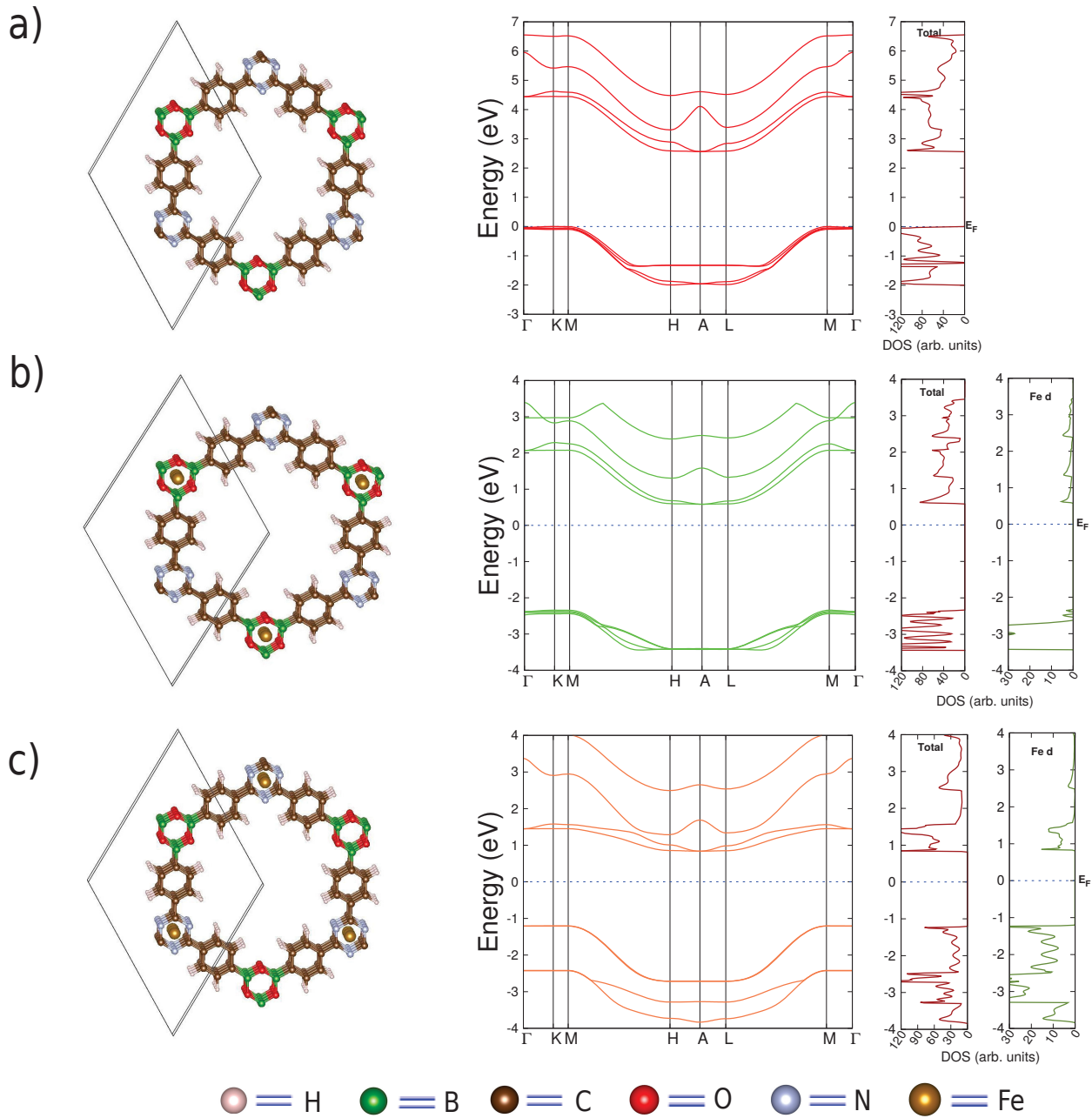

Figure 3: Optimized crystal structures (one unit cell), band structure and DOSs of a) COF-Fe-0, b) COF-Fe-1A and c) COF-Fe-1B materials. The Fe atom was intercalated in order to keep the hexagonal $P-6m2$ symmetry of the Fe-intercalated COFs.

Both the conduction and valance bands (CB and VB) were moved down by adding more Fe atoms in the COF-Fe-1A and COF-Fe-1B. The unit cells and the electronic properties of both the COF-Fe-2 and COF-Fe-3 are shown in Figure 4a-b. The band gaps of the COF-Fe-2 and COF-Fe-3 were 1.97 eV and 1.15 eV, respectively, as shown in Table 2. The band structure calculations show that two conduction bands slightly touch each other at around 0.15 eV above the Fermi energy level ($E_F$) in the $M-H-A-L-M$ direction accompanied by a large electron density reflected in the total DOSs computation as shown in Figure 4b. The large electron density around the $E_F$ is due largely to the $d$-subshell electrons of the Fe atoms as depicted in the $d$-subshell DOSs calculations. Thus, the present study reveals that the Fe atoms changed the band structure of the COFs depending on the intercalation of Fe atoms. The Fe atoms caused the COFs to become semiconductors, pushed down the conduction bands, and caused the CB to move toward the Fermi energy level.

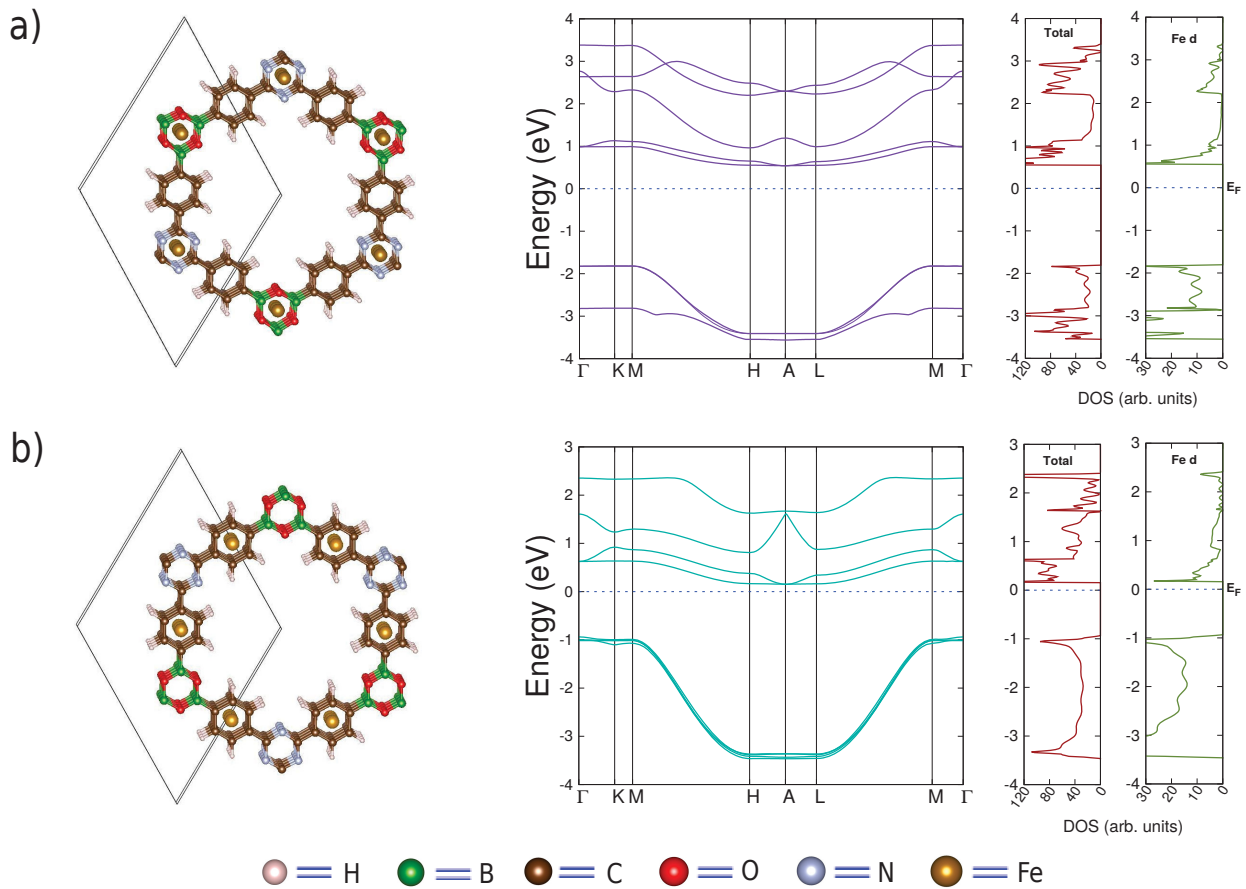

Figure 4: Optimized crystal structures (one unit cell), band structure and DOSs of a) COF-Fe-2 and b) COF-Fe-3. The Fe atom was intercalated in order to keep the hexagonal $P-6m2$ symmetry of the Fe-intercalated COFs.

Similarly, we expanded our study to include the electronic or material properties of COF-Fe-4A and COF-Fe-4B. They showed a similar type of behavior to that of COF-Fe-3, as depicted in the band structure and total DOSs calculations in Figure 5a-b. The Fe atoms pushed the conduction bands toward the Fermi level. Large electron densities were found around the $E_F$. These were caused by the Fe $d$-subshell electrons as shown in the DOSs calculations, denoted by **Fe d** in Figure 5a-b. One can clearly see that the general features and characteristic peaks of the total DOSs of both the COF-Fe-4A and COF-Fe-4B changed relative to COF-Fe-0, COF-Fe-1A, COF-Fe-1B, COF-Fe-2 and COF-Fe-3. After adding the fourth Fe atom in the COF-Fe-3 to create COF-Fe-4A and COF-Fe-4B, the CB moved slightly away from the $E_F$, which caused a lack of contact between the CB at the $E_F$ level. This can be explained by the interaction of the $d$-subshell electrons of the Fe atoms with the electron clouds of the boroxine ring in COF-Fe-4A. The shapes of the valence band (VB) and CB are similar to those of the other COFs. COF-Fe-4B shows that the CBs are closer to the $E_F$ than COF-Fe-1A, which resulted in a large peak of DOSs around the Fermi level. The present computation shows that the COF-Fe-4A had an indirect band gap of about 1.38 eV (see Table 2); thus, this material is a semiconductor similar to COF-Fe-3. COF-Fe-4B also had an indirect band gap of around 1.12 eV which was about 0.26 eV lower than COF-Fe-4A. Thus the COF-Fe-4B was also an indirect band gap semiconductor. In the case of COF-Fe-4A, where Fe was placed at the centroid of the boroxine ring, the band gap slightly increased compared to COF-Fe-4B. This is further evidence of the lower aromaticity of the boroxine ring relative to the benzene and triazine rings. The binding energy calculations (Table 2) and total energy calculations (see Supporting Information) also show that Fe intercalation at the centroid of boroxine rings is less stable than triazine rings; thus, COF-Fe-4B is more energetically stable than COF-Fe-4A.

The shape of the CB structure was slightly changed in the COF-Fe-5 due to the addition of the 5th Fe atom to COF-Fe-4A or COF-Fe-4B at the centroid of the triazine or boroxine ring. The shape of the VB was similar to the COF-Fe-3, COF-Fe-4A, and COF-Fe-4B as shown in Figure 5c. The band structure of COF-Fe-5 showed that one conduction band is overlapped with the Fermi energy level in the $\Gamma - K - M$ direction creating large electron density around the $E_F$, which was reflected in the total DOSs as shown in Figure 5c. The peaks around the $E_F$ level in the partial DOSs show that the $d$-subshell electrons of the Fe atoms provided most of the large electron density. The $d$-subshell electrons of Fe atoms were interacting with the $p$-subshell electrons of the C in the triazine ring, resulting in a conducting region around this ring. This study shows that triazine rings in COF-Fe-5 help to increase the conductivity character, just like the Fe between benzene rings did in COF-Fe-3. To put it another way, the Fe between the triazine and benzene rings on these COFs facilitated electrons passing through the rings perpendicular to the basal plane of this material, i.e. from one layer to the next layer. COF-Fe-5 had a small direct band gap of around 1.18 eV (see Table 2), resulting in a semiconductor. This value was well below the values for the pristine 3D-COFs calculated by Lukose et al. (2.3 - 4.2 eV).[36]

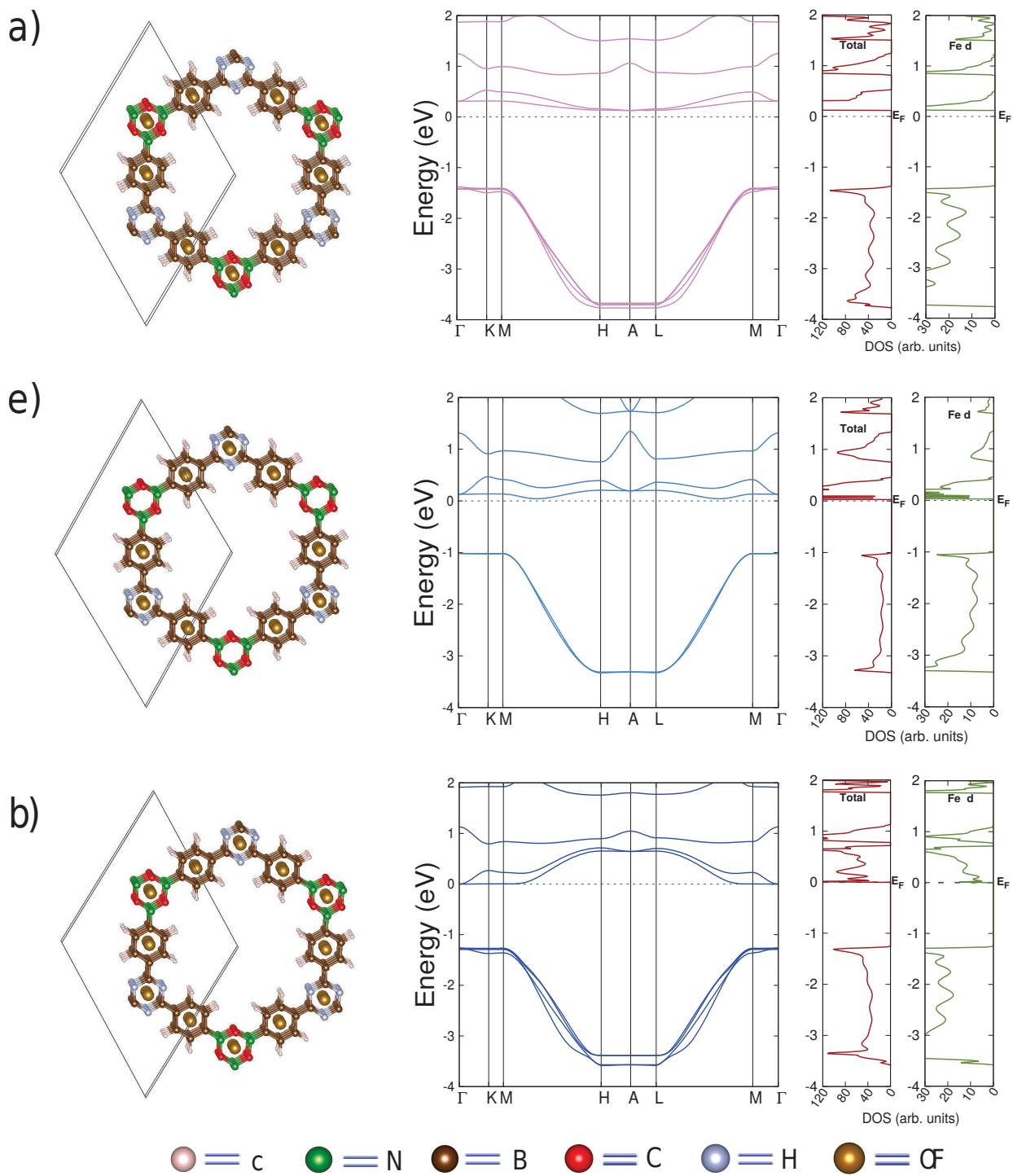

Figure 5: Optimized crystal structures (one unit cell), band structure, and DOSs of a) COF-Fe-4A, b) COF-Fe-4B and c) COF-Fe-5. The Fe atom was intercalated in order to keep the hexagonal $P-6m2$ symmetry in the Fe-intercalated COFs. The contribution of $d$-subshell electrons of Fe in total DOSs was shown in the right hand side of the figure and it was highlighted by "Fe d".

The band structure and DOSs calculations indicate that the Fe-intercalated COF materials have electronic properties similar to well known thermoelectric materials, such as $La_{3-y}X_4$ (X = S, Se, Te),[37] with band gap energies between ∼1.23 - 2.06 eV. The present calculations found that the band gap of Fe-intercalated COFs are in that range: COF-Fe-3, COF-Fe-4A, COF-Fe-4B, and COF-Fe-5 have band gaps of 1.15, 1.38, 1.12, and 1.18 eV, respectively (see Table 2). Additionally, the Fe-intercalated COFs studied are semiconductors, and all thermoelectric materials are, in general, semiconductors.[38,39] Conceptually, to obtain a good thermoelectric material, both the Seebeck coefficient and electrical conductivity of the materials must be optimal.[40] The Seebeck coefficient is directly proportional to the effective mass of the electrons ($m_e^*$) in the crystal, and the electrical conductivity ($\sigma$) is inversely proportional to $m_e^*$. The band structures of the intercalated-COFs studied here (COF-Fe-3, COF-Fe-4A, COF-Fe-4B, and COF-Fe-5 among all) show that these Fe-intercalated COF materials should have an optimal value of the $m_e^*$ in the $M - H$ and $L - M$ region. The effective mass of carriers in the aforementioned intercalated-COFs were calculated and reported in Table 2. The effective mass is related to the curvature of the band at the top of the valence band or at the bottom of the conduction band. The method developed by LeBahers et al.[41] was used to compute the $m_e^*$ of the carriers.

It has been observed that the Seebeck coefficient can be enhanced through a distortion of the electronic DOSs by doping or adding impurities to the system,[40,42] resulting in the reduction of the thermal conductivity. Such situations can occur when the VB or CB of the materials resonates with the localized impurity energy level. The present computations show that the band and DOSs structures of six intercalated-COFs (COF-Fe-1B, COF-Fe-2, COF-Fe-3, COF-Fe-4A, COF-Fe-4B and COF-Fe-5) are distorted around the $E_F$ by adding Fe atoms to the systems, and the $E_F$ is aligned with a large and narrow peak of the DOSs, resulting in an enhanced Seebeck coefficient of the aforementioned materials studied here. The $d$-subshell electron density is the main contributing component in the total DOSs, which results in the band and total DOSs distortion around the $E_F$. Thus, the effect of this distortion in the total DOSs changed the effective mass and produced a higher Seebeck coefficient as proposed by Snyder and co-workers.[40,42] These six materials show the similar type of band and DOSs structures while being porous, as well as a small indirect band gap (except COF-Fe-5 which has a direct band gap of 1.18 eV) which indicates a new type of porous thermoelectric materials. If the approximation used here holds, an electron could move from the valence band to the conduction band at room temperature, and there would be no need to use extra energy to excite or promote the electrons. The surface area of these materials is on the order of $10^3$ m$^2$/g, which makes them potentially more sensitive to small heat or electrical changes. This means that if these six materials can be fabricated experimentally, they may show thermoelectric behavior. Our present computations also show that the threshold adsorption wavelengths of these materials COF-Fe-3, COF-Fe-4A, COF-Fe-4B and COF-Fe-5, are in the range of 827 - 1035 nm, which is located in the near infrared (IR) region.

The purpose of this paper is to suggest the synthesis and predict the properties of these Fe-intercalated COFs, and while none of the materials (COF-Fe-0, COF-Fe-1A, COF-Fe-1B, COF-Fe-2, COF-Fe-3, COF-Fe-4A, COF-Fe-4B and COF-Fe-5) studied here have been synthesized experimentally. However, the calculated vibrational frequencies and binding energies predict that theses materials are stable. The frequency analysis and binding energy has been shown enough to demonstrate a material stability and a posteriori synthesis in previous work.[11,22] Thus, we expect Fe-intercalated COFs to be readily fabricated experimentally by Fe intercalation between weakly bound pristine COF layers.

Although Ca-intercalated COFs were reported in the past, the base materials and linkers are different than those in the current work. In the Ca-intercalated COFs, Meng and co-workers[14] used benzene and boroxine linkers. More specifically, they used benzene diboronic acid as an organic linker and they proposed that Ca-intercalated COFs could be useful for $H_2$ storage materials. In the current Fe-intercalated COFs, we used triazine organic linkers in the design of the new materials. Additionally, our new materials are not intended for $H_2$ storage, but rather for increasing the conductivity and reducing the band gap in COFs. Here we proposed that these Fe-intercalated COFs may be useful for small gap materials, which is an important property of crystalline porous materials. We designed these compounds with the triazine included because it has $\pi$ electrons to facilitate electron diffusion, while the boroxine ring does not have $\pi$ electrons. We imagined that the intercalation of transition metal atoms (here Fe) would also improve the conductivity of COFs. These two elements (triazine and Fe) form a unique idea which can be used in semi-conducting technology as a way to improve conductivity and stability in nanoporous materials in future applications.

# CONCLUSIONS

In summary, the structure and electronic properties of a new kind of Fe-intercalated COF materials were studied using dispersion-corrected unrestricted hybrid DFT methods. The results for the new proposed intercalated-COF materials are quite interesting and this Fe-intercalation approach is a promising way to tune the band gap from big to small while keeping the original symmetry of the pristine COF. Our findings suggest a new type of porous materials having entirely different electronic properties compared to usual porous materials such as pristine COFs or MOFs. The present study also shows how the insulating pristine COF converts to a direct band gap semiconductor when COF-Fe-0 is intercalated by five Fe atoms. Specifically, we focused on the fundamental properties required from a semiconductor to make it usable as a thermoelectric, such as band structures, DOSs, bandgap, and effective electron mass. The intercalation of Fe between layers in COFs showed that the band convergence and DOSs distortion could enhance the Seebeck coefficient. The main contributing component in the total DOSs was investigated by computing the $d$-subshell electron density of Fe atoms. The four Fe-intercalated COFs (COF-Fe-3, COF-Fe-4A, COF-Fe-4B and COF-Fe-5) can be characterized as thermoelectric porous materials, since the $d$-subshell electron density of Fe atoms distorted the band and total DOSs, resulting in a large peak around the $E_F$. These designed intercalated-COFs have an unusual band structure with band gaps around 1.12-1.97 eV, making them semiconductors at room temperature. In comparison, typical pristine COFs often have very large band gaps of around 2.3-4.2 eV. This approach broadens the scope and complexity of nanoporous COF materials, and opens the door to a more informed search for new electronic organic materials that combine the features of a porous framework structure with tunable electronic properties which are suitable for semiconducting applications. Our work indicates that Fe intercalation is a promising approach for converting an insulating pristine COF to semiconducting porous materials for device applications based on its electronic properties. Understanding the new properties and behavior of these intercalated-COF materials will allow us to design new COFs that are intercalated by other transition elements in the periodic table for specific potential new applications in the near future. Further computational investigation and study will focus on the effect of the other transition metals on the electronic properties of other designed COF materials.

## Acknowledgement

S.P. and J.L.M-C. were supported by Florida State University (FSU). J.L.M-C. gratefully acknowledges the support from the Energy and Materials Initiative at FSU. S. P. is grateful to Ms. Stephanie Marxsen, Dr. Yohanes Pramudya and Mr. Oluwagbenga Iyiola from the Department of Chemical Engineering at FSU for helpful discussions and guidance with computational resources. The authors thank the High Performance Computer cluster at the Research Computing Center (RCC) in FSU for providing computational resources and support. The authors thank the reviewers for their constructive comments and suggestions.

*Supporting Information*: Individual spins of Fe atoms; the optimized crystallographic information (.cif) files of the pristine and intercalated COF materials and detailed structural information of COF-Fe-0, COF-Fe-1A, COF-Fe-1B, COF-Fe-2, COF-Fe-3, COF-Fe-4A, COF-Fe-4B and COF-Fe-5; the total energies of these materials.

TABLE OF CONTENTS GRAPHIC

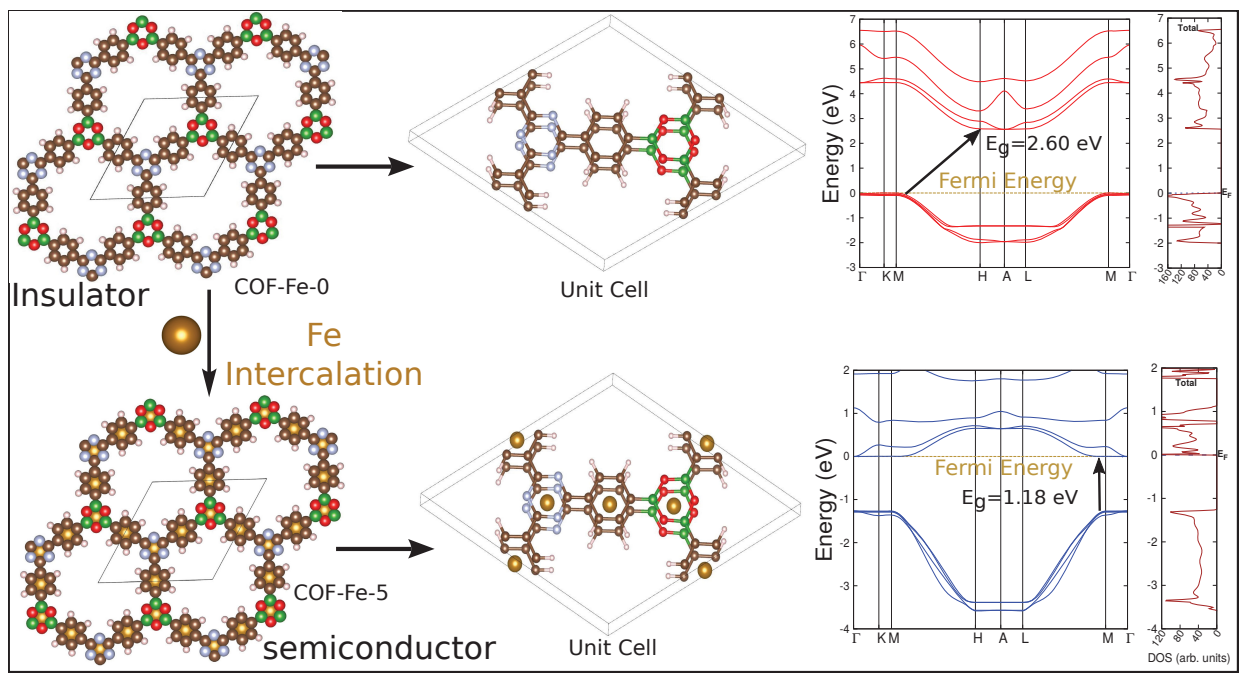

# Supplementary Information

## Iron Intercalation in Covalent-Organic Frameworks: A Promising Approach for Semiconductors

Srimanta Pakhira[1,2], Kevin P. Lucht[1,2], Jose L. Mendoza-Cortes[1,2,*]

[1] Department of Chemical & Biomedical Engineering, Florida A&M University - Florida State University, Joint College of Engineering, Tallahassee, Florida, 32310, USA.

[2] Condensed Matter Theory, National High Magnetic Field Laboratory (NHMFL), Scientific Computing Department, Materials Science and Engineering, High Performance Materials Institute (HPMI), Florida State University, Tallahassee, Florida, 32310, USA..

E-mail: mendoza@eng.famu.fsu.edu

# Contents

# 1 Individual spins of the Fe atoms in the intercalated COFs

We used the dispersion-corrected unrestricted B3LYP-D hybrid Density Functional Theory (DFT) method in this work, i.e. UB3LYP-D.[1–9] All the computations were performed using CRYSTAL14 program NOT VASP as stated in the main manuscript. This means that we have used Gaussian basis sets not plane-wave basis sets. Thus, we used the unrestricted wave function to consider the spin polarization of Fe atoms. To specify the spin unrestricted calculations, we used "SPIN" keyword. The spins of Fe atoms were fixed by the "ATOMSPIN" and "SPINLOCK" keywords in the calculations using the aforementioned program/code. In order to converge the Self-Consistent Field (SCF) calculations, we assigned the spins of the individual Fe atoms in the intercalated COFs. Initially, we locked the spins of Fe atoms up to four SCF cycles, and after that, the spins of all atoms were relaxed. We have also performed Mulliken spin population analysis in the present computations. We found that C, H, N, B, O have no unpaired spins and Fe atoms have unpaired spins. Table S1 shows the individual spins of Fe and average spin of Fe. The individual spins of the iron atoms (Fe) and average spin of Fe in the intercalated COFs (COF-Fe-0, COF-Fe-1A, COF-Fe-1B, COF-Fe-2, COF-Fe-3, COF-Fe-4A, COF-Fe-4B and COF-Fe-5) are shown in Table S1.

Table S1: Individual spins of the Fe atoms in COF-Fe-0, COF-Fe-1A, COF-Fe-1B, COF-Fe-2, COF-Fe-3, COF-Fe-4A, COF-Fe-4B and COF-Fe-5.

| COFs      | No of Fe | Fe1   | Fe2   | Fe3   | Fe4   | Fe5   | Average Spin of Fe |
|-----------|----------|-------|-------|-------|-------|-------|--------------------|
| COF-Fe-0  | 0        | N/A   | N/A   | N/A   | N/A   | N/A   | N/A                |
| COF-Fe-1A | 1        | 2.373 | N/A   | N/A   | N/A   | N/A   | 2.373              |
| COF-Fe-1B | 1        | N/A   | 2.334 | N/A   | N/A   | N/A   | 2.334              |
| COF-Fe-2  | 2        | 2.363 | 3.283 | N/A   | N/A   | N/A   | 2.826              |
| COF-Fe-3  | 3        | N/A   | N/A   | 2.635 | 2.635 | 2.635 | 2.635              |
| COF-Fe-4A | 4        | 2.129 | N/A   | 2.624 | 2.624 | 2.624 | 2.500              |
| COF-Fe-4B | 4        | N/A   | 2.878 | 2.606 | 2.606 | 2.606 | 2.674              |
| COF-Fe-5  | 5        | 2.326 | 2.876 | 2.610 | 2.610 | 2.610 | 2.606              |

# 2 Optimized Structures (.cif format)

The optimized structures are provided below in .cif format.

## 2.1 COF-Fe-0: Pristine COF

**Total Energy** $E_{COF-Fe-0}$ = **-1271.9764747 a.u.**

```
data_COF-Fe-0
_symmetry_space_group_name_H-M    'P-6M2'
_symmetry_Int_Tables_number       187
_symmetry_cell_setting            hexagonal
loop_
_symmetry_equiv_pos_as_xyz
  x,y,z
  -y,x-y,z
  -x+y,-x,z
  x,y,-z
  -y,x-y,-z
  -x+y,-x,-z
  -y,-x,z
  -x+y,y,z
  x,x-y,z
  -y,-x,-z
```

```
  -x+y,y,-z
  x,x-y,-z
_cell_length_a                      14.7275
_cell_length_b                      14.7275
_cell_length_c                      3.2408
_cell_angle_alpha                   90.0000
_cell_angle_beta                    90.0000
_cell_angle_gamma                   120.0000
loop_
_atom_site_label
_atom_site_type_symbol
_atom_site_fract_x
_atom_site_fract_y
_atom_site_fract_z
_atom_site_U_iso_or_equiv
_atom_site_adp_type
_atom_site_occupancy
C001   C    -0.38821   -0.44832    0.00000    0.00000  Uiso   1.00
C007   C     0.04784   -0.39468   -0.00000    0.00000  Uiso   1.00
H013   H     0.40727    0.10371   -0.00000    0.00000  Uiso   1.00
H019   H    -0.40041   -0.30980    0.00000    0.00000  Uiso   1.00
C025   C    -0.38476    0.38476   -0.00000    0.00000  Uiso   1.00
C028   C    -0.44247    0.44247    0.00000    0.00000  Uiso   1.00
B031   B     0.38811   -0.38811   -0.00000    0.00000  Uiso   1.00
N034   N    -0.27996    0.27996   -0.00000    0.00000  Uiso   1.00
C037   C     0.44818   -0.44818   -0.00000    0.00000  Uiso   1.00
O040   O     0.27968   -0.27968    0.00000    0.00000  Uiso   1.00
```

## 2.2 COF-Fe-1A: Fe-intercalated COF

**Total Energy** $E_{COF-Fe-0}$ = -2535.4864722 a.u.

```
data_COF-Fe-1A
_symmetry_space_group_name_H-M      'P-6M2'
_symmetry_Int_Tables_number         187
_symmetry_cell_setting              hexagonal
loop_
_symmetry_equiv_pos_as_xyz
  x,y,z
  -y,x-y,z
  -x+y,-x,z
  x,y,-z
  -y,x-y,-z
  -x+y,-x,-z
  -y,-x,z
  -x+y,y,z
  x,x-y,z
  -y,-x,-z
  -x+y,y,-z
  x,x-y,-z
_cell_length_a                      14.8551
_cell_length_b                      14.8551
_cell_length_c                      3.3120
_cell_angle_alpha                   90.0000
```

```
_cell_angle_beta                    90.0000
_cell_angle_gamma                   120.0000
loop_
_atom_site_label
_atom_site_type_symbol
_atom_site_fract_x
_atom_site_fract_y
_atom_site_fract_z
_atom_site_U_iso_or_equiv
_atom_site_adp_type
_atom_site_occupancy
C001   C    -0.38781   -0.45017    0.00000   0.00000  Uiso  1.00
C007   C     0.04457   -0.39692    0.00000   0.00000  Uiso  1.00
H013   H     0.40944    0.10555   -0.00000   0.00000  Uiso  1.00
H019   H    -0.39947   -0.31275    0.00000   0.00000  Uiso  1.00
C025   C    -0.38433    0.38433   -0.00000   0.00000  Uiso  1.00
C028   C    -0.44148    0.44148   -0.00000   0.00000  Uiso  1.00
B031   B     0.39123   -0.39123    0.00000   0.00000  Uiso  1.00
N034   N    -0.28036    0.28036   -0.00000   0.00000  Uiso  1.00
C037   C     0.44999   -0.44999    0.00000   0.00000  Uiso  1.00
O040   O     0.27929   -0.27929   -0.00000   0.00000  Uiso  1.00
FE043  Fe    0.33333   -0.33333    0.50000   0.00000  Uiso  1.00
```

## 2.3  COF-Fe-1B: Fe-intercalated COF

**Total Energy** $E_{COF-Fe-0}$ = -2535.5437897 a.u.

```
data_COF-Fe-1B
_symmetry_space_group_name_H-M      'P-6M2'
_symmetry_Int_Tables_number         187
_symmetry_cell_setting              hexagonal
loop_
_symmetry_equiv_pos_as_xyz
  x,y,z
  -y,x-y,z
  -x+y,-x,z
  x,y,-z
  -y,x-y,-z
  -x+y,-x,-z
  -y,-x,z
  -x+y,y,z
  x,x-y,z
  -y,-x,-z
  -x+y,y,-z
  x,x-y,-z
_cell_length_a                      14.8000
_cell_length_b                      14.8000
_cell_length_c                      3.2922
_cell_angle_alpha                   90.0000
_cell_angle_beta                    90.0000
_cell_angle_gamma                   120.0000
loop_
_atom_site_label
_atom_site_type_symbol
```

```
_atom_site_fract_x
_atom_site_fract_y
_atom_site_fract_z
_atom_site_U_iso_or_equiv
_atom_site_adp_type
_atom_site_occupancy
C001   C    -0.38947   -0.44767    0.00000   0.00000   Uiso   1.00
C007   C     0.04916   -0.39445   -0.00000   0.00000   Uiso   1.00
H013   H     0.40661    0.10137   -0.00000   0.00000   Uiso   1.00
H019   H    -0.40166   -0.30998    0.00000   0.00000   Uiso   1.00
C025   C    -0.38663    0.38663    0.00000   0.00000   Uiso   1.00
C028   C    -0.44354    0.44354    0.00000   0.00000   Uiso   1.00
B031   B     0.38785   -0.38785    0.00000   0.00000   Uiso   1.00
N034   N    -0.27862    0.27862    0.00000   0.00000   Uiso   1.00
C037   C     0.44758   -0.44758   -0.00000   0.00000   Uiso   1.00
O040   O     0.27992   -0.27992    0.00000   0.00000   Uiso   1.00
FE043  Fe   -0.33333    0.33333    0.50000   0.00000   Uiso   1.00
```

## 2.4 COF-Fe-2: Fe-intercalated COF

**Total Energy** $E_{COF-Fe-0}$ = **-3799.0160587 a.u.**

```
data_COF-Fe-2
_symmetry_space_group_name_H-M    'P-6M2'
_symmetry_Int_Tables_number       187
_symmetry_cell_setting            hexagonal
loop_
_symmetry_equiv_pos_as_xyz
  x,y,z
  -y,x-y,z
  -x+y,-x,z
  x,y,-z
  -y,x-y,-z
  -x+y,-x,-z
  -y,-x,z
  -x+y,y,z
  x,x-y,z
  -y,-x,-z
  -x+y,y,-z
  x,x-y,-z
_cell_length_a                    14.9492
_cell_length_b                    14.9492
_cell_length_c                    3.3717
_cell_angle_alpha                 90.0000
_cell_angle_beta                  90.0000
_cell_angle_gamma                 120.0000
loop_
_atom_site_label
_atom_site_type_symbol
_atom_site_fract_x
_atom_site_fract_y
_atom_site_fract_z
_atom_site_U_iso_or_equiv
_atom_site_adp_type
```

```
_atom_site_occupancy
C001   C   -0.38942   -0.44890    0.00000    0.00000   Uiso   1.00
C007   C    0.04650   -0.39645   -0.00000    0.00000   Uiso   1.00
H013   H    0.40801    0.10201   -0.00000    0.00000   Uiso   1.00
H019   H   -0.40122   -0.31280    0.00000    0.00000   Uiso   1.00
C025   C   -0.38737    0.38737    0.00000    0.00000   Uiso   1.00
C028   C   -0.44285    0.44285    0.00000    0.00000   Uiso   1.00
B031   B    0.39062   -0.39062   -0.00000    0.00000   Uiso   1.00
N034   N   -0.27840    0.27840    0.00000    0.00000   Uiso   1.00
C037   C    0.44902   -0.44902   -0.00000    0.00000   Uiso   1.00
O040   O    0.27966   -0.27966   -0.00000    0.00000   Uiso   1.00
FE043  Fe   0.33333   -0.33333   -0.50000    0.00000   Uiso   1.00
FE044  Fe  -0.33333    0.33333   -0.50000    0.00000   Uiso   1.00
```

## 2.5 COF-Fe-3: Fe-intercalated COF

**Total Energy** $E_{COF-Fe-0}$ = -5062.8211239 a.u.

```
data_COF-Fe-3
_symmetry_space_group_name_H-M    'P-6M2'
_symmetry_Int_Tables_number       187
_symmetry_cell_setting            hexagonal
loop_
_symmetry_equiv_pos_as_xyz
  x,y,z
  -y,x-y,z
  -x+y,-x,z
  x,y,-z
  -y,x-y,-z
  -x+y,-x,-z
  -y,-x,z
  -x+y,y,z
  x,x-y,z
  -y,-x,-z
  -x+y,y,-z
  x,x-y,-z
_cell_length_a                    14.8311
_cell_length_b                    14.8311
_cell_length_c                    3.4436
_cell_angle_alpha                 90.0000
_cell_angle_beta                  90.0000
_cell_angle_gamma                 120.0000
loop_
_atom_site_label
_atom_site_type_symbol
_atom_site_fract_x
_atom_site_fract_y
_atom_site_fract_z
_atom_site_U_iso_or_equiv
_atom_site_adp_type
_atom_site_occupancy
C001   C   -0.38597   -0.44685    0.00000    0.00000   Uiso   1.00
C007   C    0.04847   -0.39240    0.00000    0.00000   Uiso   1.00
H013   H    0.40659    0.10439   -0.00000    0.00000   Uiso   1.00
```

```
H019   H    -0.39936  -0.30842   0.00000   0.00000  Uiso  1.00
C025   C    -0.38441   0.38441  -0.00000   0.00000  Uiso  1.00
C028   C    -0.44073   0.44073  -0.00000   0.00000  Uiso  1.00
B031   B     0.38771  -0.38771   0.00000   0.00000  Uiso  1.00
N034   N    -0.28000   0.28000  -0.00000   0.00000  Uiso  1.00
C037   C     0.44649  -0.44649   0.00000   0.00000  Uiso  1.00
O040   O     0.27983  -0.27983   0.00000   0.00000  Uiso  1.00
FE043  Fe   -0.49763   0.49763  -0.50000   0.00000  Uiso  1.00
```

## 2.6 COF-Fe-4A: Fe-intercalated COF

**Total Energy** $E_{COF-Fe-0}$ = -6326.3252772 a.u.

```
data_COF-Fe-4A
_symmetry_space_group_name_H-M    'P-6M2'
_symmetry_Int_Tables_number       187
_symmetry_cell_setting            hexagonal
loop_
_symmetry_equiv_pos_as_xyz
  x,y,z
  -y,x-y,z
  -x+y,-x,z
  x,y,-z
  -y,x-y,-z
  -x+y,-x,-z
  -y,-x,z
  -x+y,y,z
  x,x-y,z
  -y,-x,-z
  -x+y,y,-z
  x,x-y,-z
_cell_length_a                    14.9482
_cell_length_b                    14.9482
_cell_length_c                    3.4482
_cell_angle_alpha                 90.0000
_cell_angle_beta                  90.0000
_cell_angle_gamma                 120.0000
loop_
_atom_site_label
_atom_site_type_symbol
_atom_site_fract_x
_atom_site_fract_y
_atom_site_fract_z
_atom_site_U_iso_or_equiv
_atom_site_adp_type
_atom_site_occupancy
C001   C    -0.38561  -0.44844   0.00000   0.00000  Uiso  1.00
C007   C     0.04584  -0.39468   0.00000   0.00000  Uiso  1.00
H013   H     0.40834   0.10584   0.00000   0.00000  Uiso  1.00
H019   H    -0.39919  -0.31132   0.00000   0.00000  Uiso  1.00
C025   C    -0.38400   0.38400   0.00000   0.00000  Uiso  1.00
C028   C    -0.43988   0.43988   0.00000   0.00000  Uiso  1.00
B031   B     0.39010  -0.39010   0.00000   0.00000  Uiso  1.00
N034   N    -0.28041   0.28041   0.00000   0.00000  Uiso  1.00
```

```
C037   C     0.44820  -0.44820   0.00000   0.00000  Uiso  1.00
O040   O     0.27953  -0.27953   0.00000   0.00000  Uiso  1.00
FE043  Fe   -0.49468   0.49468  -0.50000   0.00000  Uiso  1.00
FE046  Fe    0.33333  -0.33333  -0.50000   0.00000  Uiso  1.00
```

## 2.7  COF-Fe-4B: Fe-intercalated COF

**Total Energy** $E_{COF-Fe-0}$ = **-6326.3861850 a.u.**

```
data_COF-Fe-4B
_symmetry_space_group_name_H-M    'P-6M2'
_symmetry_Int_Tables_number       187
_symmetry_cell_setting            hexagonal
loop_
_symmetry_equiv_pos_as_xyz
  x,y,z
  -y,x-y,z
  -x+y,-x,z
  x,y,-z
  -y,x-y,-z
  -x+y,-x,-z
  -y,-x,z
  -x+y,y,z
  x,x-y,z
  -y,-x,-z
  -x+y,y,-z
  x,x-y,-z
_cell_length_a                    14.9245
_cell_length_b                    14.9245
_cell_length_c                    3.4520
_cell_angle_alpha                 90.0000
_cell_angle_beta                  90.0000
_cell_angle_gamma                 120.0000
loop_
_atom_site_label
_atom_site_type_symbol
_atom_site_fract_x
_atom_site_fract_y
_atom_site_fract_z
_atom_site_U_iso_or_equiv
_atom_site_adp_type
_atom_site_occupancy
C001   C    -0.38779  -0.44667   0.00000   0.00000  Uiso  1.00
C007   C     0.04995  -0.39220   0.00000   0.00000  Uiso  1.00
H013   H     0.40650   0.10198  -0.00000   0.00000  Uiso  1.00
H019   H    -0.40074  -0.30876   0.00000   0.00000  Uiso  1.00
C025   C    -0.38655   0.38655   0.00000   0.00000  Uiso  1.00
C028   C    -0.44193   0.44193   0.00000   0.00000  Uiso  1.00
B031   B     0.38735  -0.38735   0.00000   0.00000  Uiso  1.00
N034   N    -0.27833   0.27833   0.00000   0.00000  Uiso  1.00
C037   C     0.44580  -0.44579  -0.00000   0.00000  Uiso  1.00
O040   O     0.28013  -0.28013   0.00000   0.00000  Uiso  1.00
FE043  Fe    0.49969  -0.49969   0.50000   0.00000  Uiso  1.00
FE046  Fe   -0.33333   0.33333   0.50000   0.00000  Uiso  1.00
```

## 2.8  COF-Fe-5: Fe-intercalated COF

**Total Energy** $E_{COF-Fe-0}$ = **-7589.8886945 a.u.**

```
data_COF-Fe-5
_symmetry_space_group_name_H-M    'P-6M2'
_symmetry_Int_Tables_number       187
_symmetry_cell_setting            hexagonal
loop_
_symmetry_equiv_pos_as_xyz
  x,y,z
  -y,x-y,z
  -x+y,-x,z
  x,y,-z
  -y,x-y,-z
  -x+y,-x,-z
  -y,-x,z
  -x+y,y,z
  x,x-y,z
  -y,-x,-z
  -x+y,y,-z
  x,x-y,-z
_cell_length_a                    15.0355
_cell_length_b                    15.0355
_cell_length_c                    3.4523
_cell_angle_alpha                 90.0000
_cell_angle_beta                  90.0000
_cell_angle_gamma                 120.0000
loop_
_atom_site_label
_atom_site_type_symbol
_atom_site_fract_x
_atom_site_fract_y
_atom_site_fract_z
_atom_site_U_iso_or_equiv
_atom_site_adp_type
_atom_site_occupancy
C001   C    -0.38729  -0.44814   0.00000   0.00000  Uiso   1.00
C007   C     0.04722  -0.39433   0.00000   0.00000  Uiso   1.00
H013   H     0.40828   0.10364  -0.00000   0.00000  Uiso   1.00
H019   H    -0.40029  -0.31146   0.00000   0.00000  Uiso   1.00
C025   C    -0.38613   0.38613  -0.00000   0.00000  Uiso   1.00
C028   C    -0.44109   0.44109   0.00000   0.00000  Uiso   1.00
B031   B     0.38992  -0.38992  -0.00000   0.00000  Uiso   1.00
N034   N    -0.27876   0.27876   0.00000   0.00000  Uiso   1.00
C037   C     0.44749  -0.44749  -0.00000   0.00000  Uiso   1.00
O040   O     0.27988  -0.27988   0.00000   0.00000  Uiso   1.00
FE043  Fe   -0.49749   0.49749  -0.50000   0.00000  Uiso   1.00
FE046  Fe    0.33333  -0.33333  -0.50000   0.00000  Uiso   1.00
FE047  Fe   -0.33333   0.33333  -0.50000   0.00000  Uiso   1.00
```